\documentclass[12pt]{iopart}
\usepackage{amssymb}
\usepackage{graphicx}% Include figure files
\usepackage{bm}% bold math
\def\qs{$1.5625\times10^{12} Ks^{-1}$}
\def\agn{Ag${}_{2869}$ }
\def\np{nanoparticle }
\def\qf{$1.5625\times 10^{13} Ks^{-1}$}

\def\etal{{\it et al }}

\begin{document}
\title[Atomic and electronic structure transformations ]{ Atomic and electronic structure transformations of silver nanoparticles under rapid cooling conditions}
\author{I. Lobato$^{1}$, J. Rojas$^{1,2}$, C. V. Landauro $^{2}$, J. Torres $^2$  }
\address{$^{1}$ Instituto Peruano de Energ\'{\i}a Nuclear, Av. Canad\'a 1470, Lima 41, Per\'u.}
\address{$^{2}$ Facultad de Ciencias F\'{\i}sicas, Universidad Nacional Mayor de San Marcos, P.O. box 14-0149, Lima~-~14, Per\'u.}
\ead{jrojast@unmsm.edu.pe}
\begin{abstract}
The structural evolution  and dynamics  of silver nanodrops Ag${}_{2896}$ (4.4 nm in 
diameter) during rapid cooling conditions has been studied by means of  
molecular dynamics  simulations and electronic density of state calculations. 
The interaction of silver atoms  is modeled  by a  tight-binding semiempirical 
interatomic potential proposed by Cleri and Rosato.   
The pair correlation functions and the pair analysis technique is applied to 
reveal the structural transition  in the process of  solidification.  
It is shown that  Ag nanoparticles  evolve 
into  different nanostructures  under different cooling processes. 
At a cooling rate  of $1.5625\times10^{13} Ks^{-1}$ the nanoparticles preserve an amorphous  
like structure containing a large amount of 1551 and 1541 pairs which 
correspond to the icosahedral symmetry. For a lower cooling rate ($1.5625\times10^{12} Ks^{-1}$ ), 
the nanoparticles transform into a crystal-like structure consisting mainly 
of 1421 and 1422 pairs which correspond to the fcc and hcp structures, 
respectively. The variations of the electronic density of states for the 
differently cooled nanoparticles are small but in correspondence with the 
structural changes.

%\textsl{Keywords}: Molecular dynamics, nanoparticles, structural transitions, electronic properties \\
%PACS: 61.43.Bn; 61.46.Df; 64.70.Nd; 73.22.-f 
\end{abstract}
%\submitto{Journal of Phys.: Condens Matter.}
%\submitto{}
\maketitle 

\section{Introduction}
\label{intro}

The research  in the field of nanoparticles is the 
basis for the development of nanotechnology \cite{Eberh2002,BalettoRMP}.
This is mainly due to the possibility of modifying the physical properties
of these systems through the control of the system size.
Silver nanoparticles are particularly interesting  because they have
a number of exciting potential applications in various fields including
electronics and biology \cite{Sidharta}.
It is also known that the structure of a material determines its properties.
For instance, in bulk face-centered cubic materials the formation of other 
structures is suppressed kinetically, whereas nanoparticles  of the same
materials exhibit different structural modifications such as icosahedron, 
decahedron and amorphous with a great variety of physical and chemical 
properties \cite{BalettoRMP,GafnerFSS,Duan2008}.  
Thus, in order to  understand the structure of metal nanoparticles, 
obtained from liquid phase, it is important to investigate their structural 
evolution during solidification under different conditions. 
Molecular dynamics (MD) simulations have proved to be among the most effective methods 
in the investigation (at atomic level) of the properties of nanoparticles,
which is difficult to carry out experimentally. 
In fact, the structural evolution during cooling of metallic nanoparticles
was extensively studied employing MD
\cite{YueQi2001,BalettoCPL,Nam2002,Shim2003,Chen2004,Delog2007}.
For the case of silver nanoparticles one can find some works 
about, for instance, investigations of the most stable structures 
and the melting process of small silver clusters \cite{AtisAg}, 
as well as the superheating of Ag nanowires \cite{Duan2008}.
The freezing of silver clusters and nanowires has also been studied 
recently \cite{Qi2008}. In such work the authors conclude that
the final structure of Ag clusters of $\sim$2.3 nm is a fcc polyhedron 
despite the different cooling rates employed.
In contrast, the study of the cooling rate dependence of solidification
microstructures of silver by Tian \etal \cite{Tian08} indicates that
the cooling rate plays a crucial effect on the silver structure of the
solid state. For this case, they find a critical cooling rate for
crystal forming of $\sim$1.0$\times10^{13} Ks^{-1}$.
Baletto \etal \cite{Baletto2002} analyzed the equilibrium structure and 
melting of some magic number Ag nanoparticles, concluding that for relatively 
large clusters  the fcc polyhedron is the most stable. 
However, according to experimental results of Reinhard \etal \cite{Reinhard}, 
both icosahedral and fcc structures are observed in large (up to 10 nm in 
diameter) free Ag clusters produced in an inert-gas-aggregation source. 
Hence, details of thermal stability, melting and freezing temperature for
nanosized non equilibrium systems still remain unclear. 

In this work, we investigate the structural transitions of silver 
nanoparticles during fast cooling. The structural changes have been 
simulated employing a tight-binding many-body potential. 
Furthermore, the possible changes of the electronic properties of silver 
nanoparticles at different temperatures were analyzed  by calculating the 
electronic density of states (DOS).
The paper is organized as follows. Sec. 2 is devoted to present the details of 
the MD simulations, the structural analysis, and the Hamiltonian model for 
the calculation of the DOS. The results and discussions are presented in 
Sec. 3. Some concluding remarks are provided in section 4.

\section{Model and methods}
\label{sec:model}
\subsection{Potential energy function}
The Ag nanoparticles are simulated by MD methods employing the many-body 
potentials developed by Rosato \cite{CleriRos} on the basis of the 
second-moment approximation to the tight-binding model (SMA-TB). 
In this framework, the band energy of an atom $i$ in a given position is 
proportional to the square root of the second moment of the local density of 
states.  The energy of this atom is then written as a sum of two terms: 
\begin{equation}
E_{tot}=\sum_i(E_i^{band}+E_i^{rep})
\end{equation}
where
\begin{equation}
E_i^{band}=-\lbrace\sum_{j,(r_{i,j} \leq r_c)}\xi^2 exp [-2q(\frac{r_{i,j}}{r_0}-1)]\rbrace^{1/2},
\end{equation}
with $\xi$ as an effective hopping integral, $r_{i,j}$ the distance between 
the atom $i$ and $j$, $r_c$ the cut-off radius for the interaction, $r_0$ the 
first-neighbour distance, and $q$ describes the distance dependence of the 
hopping integral. The second term in eqn. (1) is the repulsive energy of the 
Born-Mayer type:
\begin{equation}
E_i^{rep}=\sum_{j,(r_{i,j}\leq r_c)}A exp [-p(\frac{r_{i,j}}{r_0}-1)],
\end{equation}
The model parameters ($\xi, A,p,q$) are fitted to the bulk properties of the 
metal. The cut-off distance of the atomic interaction is set between the 
second and third neighbour distance. The parameters used in the simulation 
are taken from \cite{CleriRos}; i.e.  $\xi=1.178, A=0.1028, p=10.928, 
q=3.139$.

\subsection{Simulation process}
The MD simulations of a silver nanoparticle 
are carried out for a cubic box without periodic boundary conditions, so that 
the nanoparticle surface is free. 
The equations of motion are integrated in time using a 
velocity Verlet algorithm. Energy conservation with an error less than  
$1\times 10^{-3} \%$ was achieved with a time step of  6.4 fs. 
The almost spherical nanoparticle was prepared by cutting a spherical 
region of a desired radius from a big face centered cubic crystal.
We have considered silver nanoparticles of different sizes ranging from
147 atoms up to 2869 atoms, but the 
results presented here are for systems of 2869 atoms, \agn, 
(in some cases we also present results for nanoparticles of 147 
atoms, Ag${}_{147}$).
It is worth mentioning that $2869$ (and 147) belongs to the 
set of magic numbers for icosahedral symmetries \cite{BalettoRMP}.  

In order to obtain an equilibrium liquid-like state for the nanoparticle,
we start the simulation at 1500 K which is a 
temperature higher than the equilibrium melting temperature of the \agn  
nanoparticle  ($T_{melt}=975 K$). 
The system is kept at this temperature for $10^5$ time steps (640 ps). 
Six different quenching processes are then carried out. 
The overall cooling rate was controlled by changing the number of MD
steps at each run.
As our first cooling process we chose a slow one in which the system is cooled
from a liquid state at 1500 K to a temperature of 300 K 
employing $1.2\times10^{5}$ MD steps, which corresponds to a cooling rate
of \qs ($k1$ process).
In the last one, a fast cooling process, we employ 
$1.2\times10^{4}$  MD steps, which
corresponds to a cooling rate of \qf ($k2$ process).
It is worth mentioning that recently Chen \etal \cite{Chen2004} employed
the same cooling rates ($k1$, and $k2$) to study the structure and dynamics
of a gold nanoparticle of similar size (2112 atoms).
Thus, we consider these two cooling rates as our extreme cases so that
we can compare the results for both nanoparticles.
The other four cooling rates $k$ were chosen such that $k1<k<k2$.
On cooling, the temperature was decreased to room temperature (300 K) by 
steps of $\Delta T=10 K$. 
The internal energy and structural 
configurations are recorded during the simulation. 

\subsection{Structural  analysis methods}
\label{atstr}

\textbf{The pair correlation function}. The pair correlation function 
(PCF) $g(r)$ has been widely used to describe the atomic structure in 
amorphous, liquid and crystalline states.  This quantity is given 
by \cite{Still},
\begin{equation}
g(r)=\frac{\langle n_i(r,r+\Delta r)\rangle}{\rho 4\pi r^2 \Delta r } \;,
\end{equation}
where $\rho$ is the atomic density  (N/V), and 
$\langle n_i(r,r+\Delta r)\rangle$ is the average number of atoms within a 
spherical shell surrounded by $r$ and $r+\Delta r$ around an arbitrary atom.\\

\noindent
\textbf{Pair analysis technique}. The common neighbours analysis (CNA)
technique, introduced by Honeycutt and Anderson \cite{Honey}, is 
a standard tool for the interpretation of molecular dynamics simulations of 
structural transformations. The local environment of a pair of atoms is 
characterized by a set of four indexes ($i,j,k,m$).  The first one indicates 
whether the pair of atoms are closer than a given cutoff distance $r_c$. 
In the present work $r_c$ is chosen to be equal the semisum of the first and 
second nearest neighbours distance in the perfect fcc Ag lattice, that is 
$r_c$ = 0.35 nm. The second index $j$ is the number of common neighbours 
to the two atoms, and the third one is the number of bonds between the 
common neighbours. The fourth index is added to provide a unique 
correspondence between number and diagrams \cite{Honey}. 
For instance, the 1551 pairs characterize the icosahedral-like local 
structure whereas the 1421 and 1422 pairs represent the fcc-like and 
hcp-like local structures, respectively. 
Additionally, icosahedral and fcc systems with structural defects are 
characterized 
by the 1541 and 1431 pairs, respectively. Finally, it is worth
mentioning that pairs are named as type I if $i=1$ and type II otherwise.

\subsection{Electronic structure}
In order to calculate the electronic properties of silver nanoparticles
we employ a {\it tight-binding} Hamiltonian given by
\begin{eqnarray}
H & = & \sum_{\vec{R}}\sum_{\vec{R}'} |\vec{R}\rangle H_{R,R'} \langle \vec{R}'|
\label{Hamil}
\end{eqnarray}
where $\{|\vec{R}\rangle\}$ is the orthonormal (atomic-like) basis set
centered at the site $\vec{R}$ with one orbital $s$ in each
site. The matrix elements $H_{R,R'}$ are defined by
\begin{eqnarray}
H_{R,R'} & = & \left\{ \begin{array}{ll}
\varepsilon_0 & \texttt{if\ } \vec{R}=\vec{R'} \\
t_0 	& \texttt{if\ } |\vec{R}-\vec{R'}|\le r_c \\
0 	& \texttt{if\ } |\vec{R}-\vec{R'}| > r_c \\
\end{array} \right.
\end{eqnarray}
with $\varepsilon_0$ and $t_0$ as the on-site and hopping terms, respectively. 
$r_c$ is the same parameter employed in the CNA technique (see above).
The atomic positions $\{\vec{R}\}$, known from the MD
simulation (see section \ref{atstr}), are fixed for these calculations.

A physically relevant quantity for the study of the electronic properties
of the system under study is the density of states  projected on
the atomic position $R$, also known as the local density of states (LDOS). 
This quantity can be determined from
\begin{eqnarray}
n_{R}(\varepsilon) = \lim_{\gamma\to 0}\left\{-\frac{1}{\pi}Im\,G_{R,R}(\varepsilon-i\gamma)\;\right\}
\label{DOS}
\end{eqnarray}
where $G_{R,R}(\varepsilon-i\gamma)$ is the diagonal element of the 
one-particle Green function 
$G=(\varepsilon-i\gamma-H)^{-1}$. The recursion procedure is a real-space
technique to tridiagonalize a symmetric (Hermitian) Hamiltonian such that
$G$ can be expressed as a continued fraction formed by the so-called
recursion coefficients. Details about the recursion method can be found
in \cite{Haydock80}.

It is worth mentioning that this procedure allows us to collect LDOS in 
differente ways. Considering that an atom in a fcc-solid 
have 12 first neighbours, for the nanoparticle we sum all the LDOS
corresponding to atoms with 12 neighbours within $r_c$ (fcc-like DOS, FDOS).
Thus, in a well ordered nanoparticle with fcc-like structure the FDOS
will be the main contribution to the total DOS. In this way we can
compare the structural and electronic changes during the transformation.

\section{Results and discusion}
\label{results}
\subsection{Atomic structure of silver nanoparticles}

\begin{figure}{}
\centering
\includegraphics*[width=0.8\textwidth]{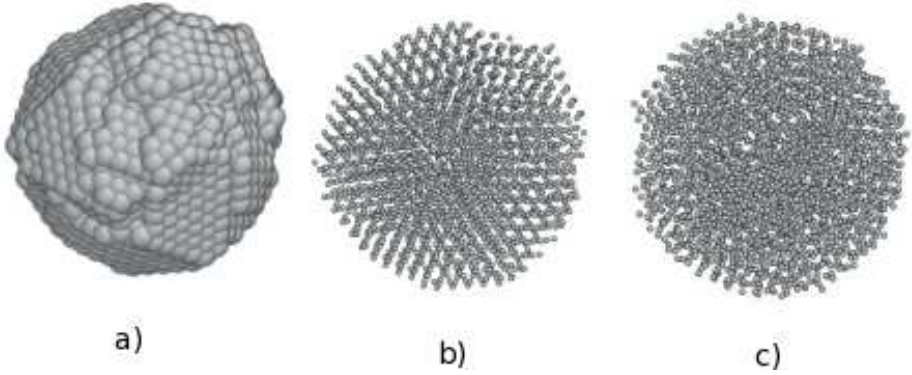}
\caption{Atomic arrangements of the silver nanoparticle \agn 
at 300 K cooled at rates of:  a),b)  \qs, and c) \qf. 
The snapshots a) and b) correspond to the same configuration but different
orientation. Visualization of atomic arrangements has been made employing
the AtomEye software \cite{eye}.}
\label{fig:fig1}
\end{figure}

The nanoparticle \agn is initially in the
liquid state forming a nanodrop of approximately 4.4 nm diameter. 
As a result of quenching at different cooling rates we obtain, at  300 K, 
nanoparticles either in  a crystal-like (cooling rate $k1=$\qs) or 
metallic glass (cooling rate $k2=$\qf) structure, 
as illustrated qualitatively in  Fig.~\ref{fig:fig1}. 
An imperfect nanocrystal with noticeable faceted morphology is obtained at a 
lower cooling rate ($k1$), as shown in Fig.~\ref{fig:fig1}(a) and (b).
At a fast cooling rate  ($k2$) nucleation of equilibrium phases are 
suppressed, and atoms exhibit random arrangements.

\begin{figure}{}
\centering
\includegraphics*[width=0.8\textwidth]{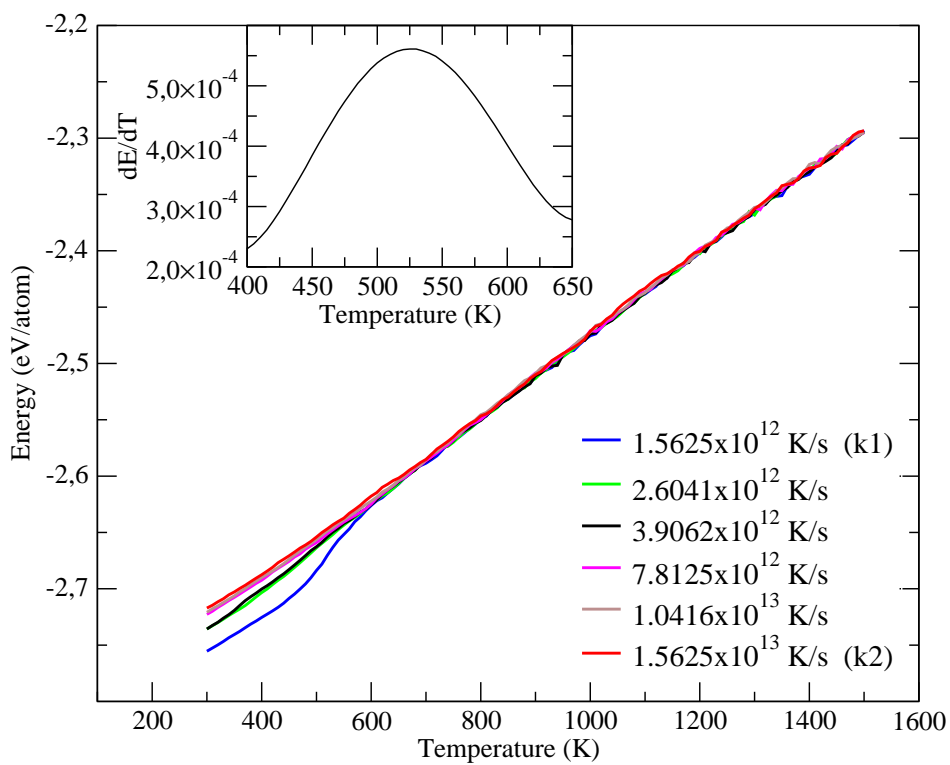}
\caption{Caloric curves  of the Ag${}_{2869}$ nanoparticle for various cooling rates.
The curve of the fitted derivative $dE/dT$ versus T for $k1=$\qs is shown
in the inset.}
\label{fig:fig2}
\end{figure}

Fig.~\ref{fig:fig2} shows the variation of the average internal energy, 
$E_t$,  as a function of temperature for various cooling rates.
At a cooling rate of $k1=$\qs the internal energy undergoes a sharp 
variation as the temperature decreases from approximately 600 to 500 K. 
As shown in the inset of Fig.~\ref{fig:fig2}, the transformation from 
liquid to crystal occurs in a wide temperature range (650 - 450 K): 
the peak of the heat capacity curve $dE/dT$, located at 527 K, is not sharp. 
On the other hand, at a cooling rate of $k2=$\qf the caloric curve $E_t$ 
has no inflexion in the whole temperature range. The continuous change of 
the curve indicates that the system does not suffer structural changes; 
i.e. an amorphous solid is obtained in the undercooling state. 
The reason for this is that high cooling rate restricts the atomic 
diffusion. 
This behavior is very similar to that reported by Chen \etal \cite{Chen2004}
for a gold  nanoparticle of 2112 atoms.
The critical cooling rate, $k_c$, is the minimum value which is necessary to 
avoid crystallization. For the nanoparticle \agn we estimate $k_c$
from the analysis of the time evolution of the energy, $E(t)$, as well as
its temperature evolution, $E(T)$ (see Fig.~\ref{fig:fig2}), and
the corresponding PCFs $g(r)$. Thus, we obtain the value of
$k_c\simeq 7.8 \times 10^{12} Ks^{-1}$. 
Although this value is close to that obtained for solid silver \cite{Tian08},
$k_c$ depends on the size of the nanoparticle.
It is also noticeable that the silver nano glass, obtained at the cooling 
rate of \qf, is very unstable.  For instance, at room temperature after 6.4 ps 
the nano glass transforms to the crystal-like structure in agreement with 
\cite{Qi2008}.
\begin{figure}{}
\centering
\includegraphics*[width=0.8\textwidth]{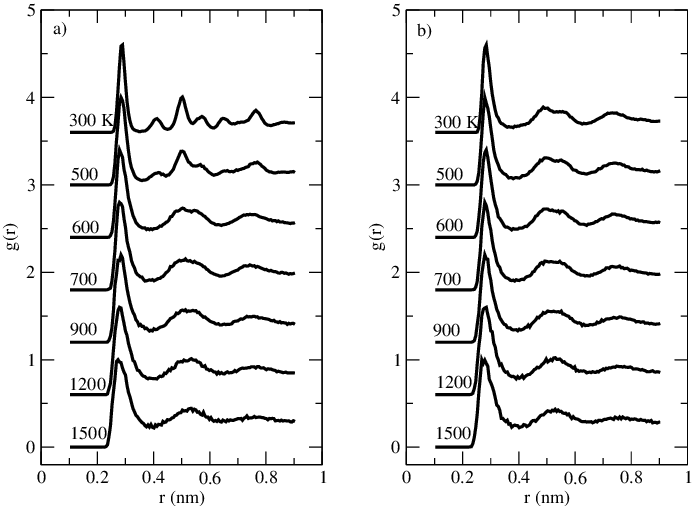}
\caption{Pair correlation functions of the Ag${}_{2869}$ nanoparticle  cooled at a 
rate of  a) \qs , b)\qf.}
\label{fig:fig3}
\end{figure}
\begin{figure}{}
\centering
\includegraphics*[width=1.0\textwidth]{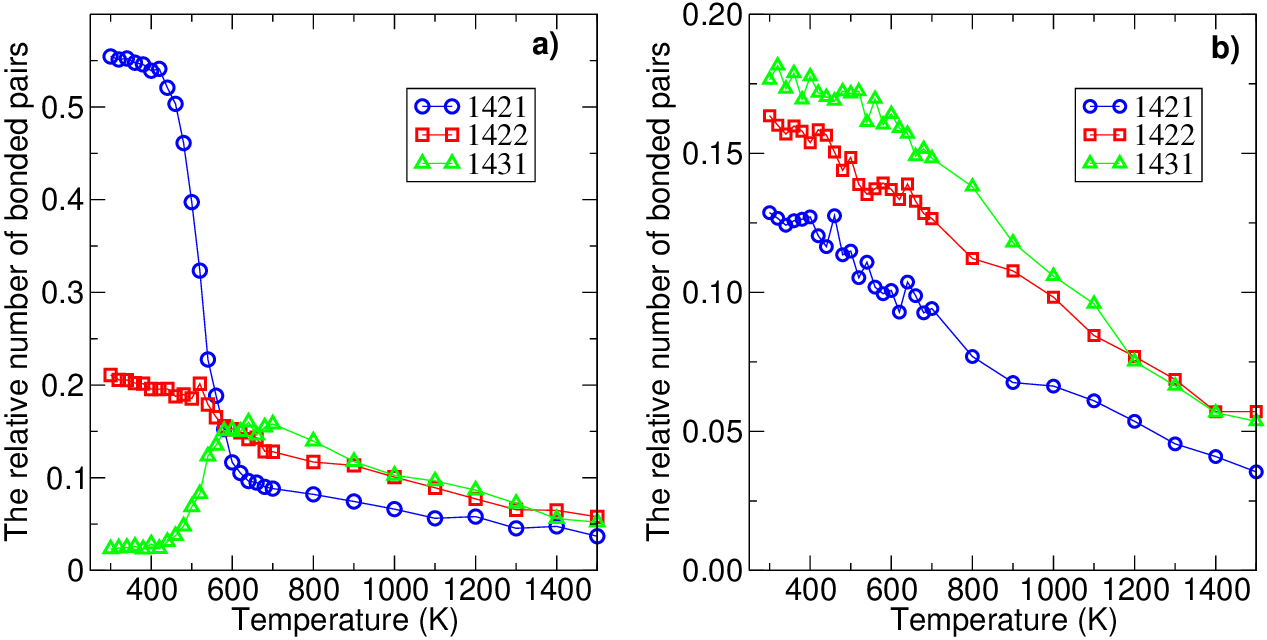}
\caption{ The relative number of bonded pairs  1421, 1422 and 1431 during 
freezing of the Ag${}_{2869}$ nanoparticle at cooling rate of: a) \qs, b) \qf}
\label{fig:fig4}
\end{figure}
\begin{figure}{}
\centering
\includegraphics*[width=1.0\textwidth]{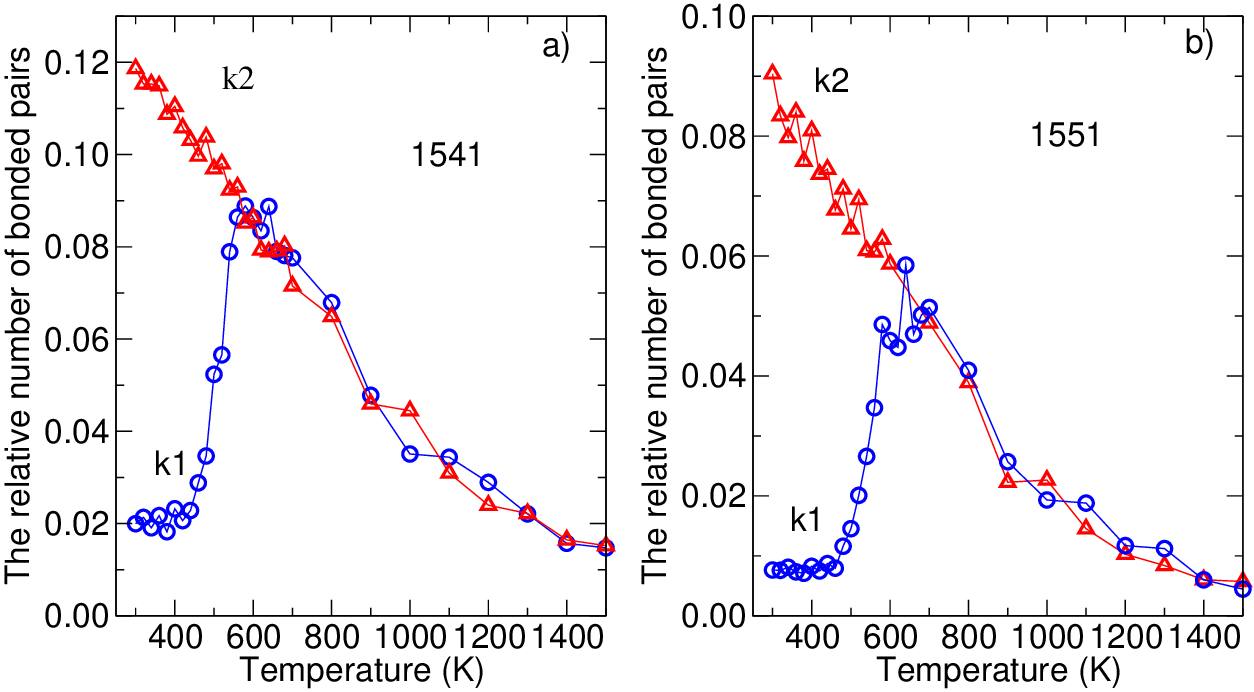}
\caption{The Relative number of bonded pairs in the \agn nanoparticle: 
a) 1541, b) 1551. k1 and k2 are the cooling rates of \qs and \qf respectively.}
\label{fig:fig5}
\end{figure}

The changes in the atomic distribution within the nanoparticles, 
during the cooling process, is extracted from the pair correlation function. 
Fig.~\ref{fig:fig3} shows the PCF at several temperatures and for the
$k1$ and $k2$ cooling rates. First, in the temperature interval from
1500 to 600 K, we note for both cooling processes that the 
PCFs are identical, and reveal the typical structural features of an 
amorphous liquid with short-range topological ordering. 
At the cooling rate of \qf the amorphous structure of the \agn nanoparticle 
is conserved down to room temperature. As temperature decreases, it is clearly 
noticeable the splitting of the second peak in the PCF, which is 
characteristic of amorphous structures \cite{Hui2002}. On the other hand, 
the PCF of the \agn nanoparticles cooled at a rate of \qs, below 500 K 
shows a typical structure corresponding to the fcc crystal.

More detailed information on the atomic structure in the \np can be obtained 
using CNA. At the beginning liquid state (1500 K), the most abundant pairs 
are 1201 (21\%), 1311 (21\%), 1101 (12\%), and 1422 (5.7\%), for type I; and 
2101, 2211 pairs, for type II. Figs.~ \ref{fig:fig4} and \ref{fig:fig5} 
show the variation of the relative numbers of several principal bonded pairs 
versus temperature. Quantities are normalized such that the total numbers of 
pairs with $i=1$ is unity.
For temperatures above 650 K the number of principal pairs decreases slowly,
but in the same proportion, after increasing the temperature which occurs
independently of the cooling proceses ($k1$ or $k2$). This is an indication
that the structure of the nanoparticle is not changing appreciably, which is
in agreement with the PCFs analysis (see Fig.~\ref{fig:fig3}).
However, for temperatures below $\sim$625 K the \agn nanoparticle, 
cooled at a rate of \qs, undergo  drastic structural changes. 
The number of 1421 pairs increases from 8\% to 50\% when
the temperature decreases from 625K to 450 K, while the number of 1431 pairs, 
which represents defect-fcc structure, practically falls to zero. Furthermore, 
the number of 1541 and 1551 bonded pairs increases equally, when temperature 
decreases to 625 K, for both $k1$ and $k2$ cooling processes. Below this 
temperature the ratio of 1541 and 1551 pairs falls practically to zero for a 
relatively slow cooling rate of \qs, while for a faster cooling process (e.g. 
$k2$), the ratio of these pairs increases, reaching values of 12\%, and 
9\% at room temperature, respectively (see figure \ref{fig:fig5}). 
It seems that a low cooling rate allows the reorganization of the atomic 
order. As the present simulation reveals, the fcc crystalline silver 
nanoparticle of 4.4 nm size is more stable than an icosahedral one, 
which is consistent with the results of Baletto \etal \cite{Baletto2002}.
These results are in good agreement with the CNA of a
gold nanoparticle of 2112 atoms \cite{Chen2004}.
In the case of the small nanoparticle of 147 atoms the number of 1422 pairs,
corresponding to hcp-like structures, increases up to 35\% (300 K) 
whereas the number of 1421 and 1551 pairs reach 
the values of $\sim10$\% at room temperatures.
In summary, for the small (147 atoms) and large (2869 atoms) nanoparticles 
predominate, after the transition, the fcc-like and hcp-like structures, 
respectively.
A more detailed study in function of the particle size is necessary in order
to identify a possible threshold between different structures, as was reported 
in the case of Ni clusters by Qi \etal \cite{YueQi2001}.

\begin{figure}{}
\centering
\includegraphics[width=0.95\textwidth]{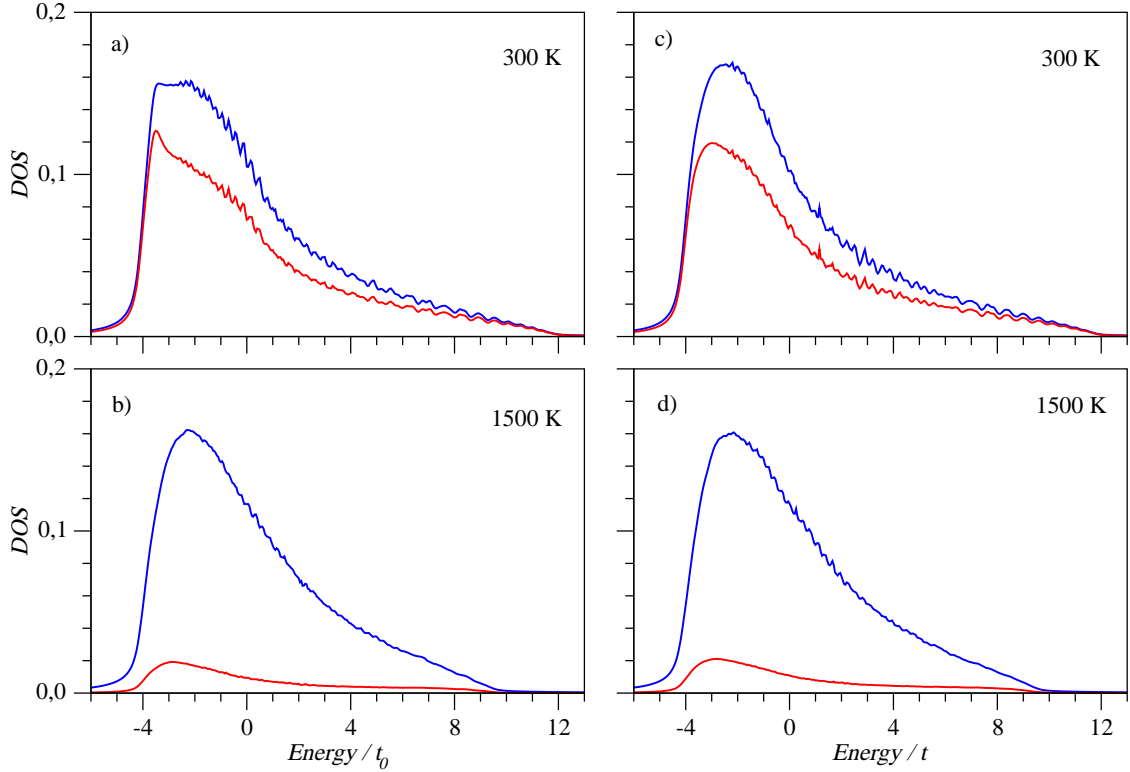}
\caption{total DOS (blue lines), and FDOS (red lines)   
of the Ag${}_{2869}$ nanoparticle for two cooling rates: $k1=$\qs (a) and b)) 
and $k2=$\qf (c) and d)). The corresponding temperatures are indicated 
in the panels. }
\label{fig:fig6}
\end{figure}

\begin{figure}{}
\centering
\includegraphics[width=0.95\textwidth]{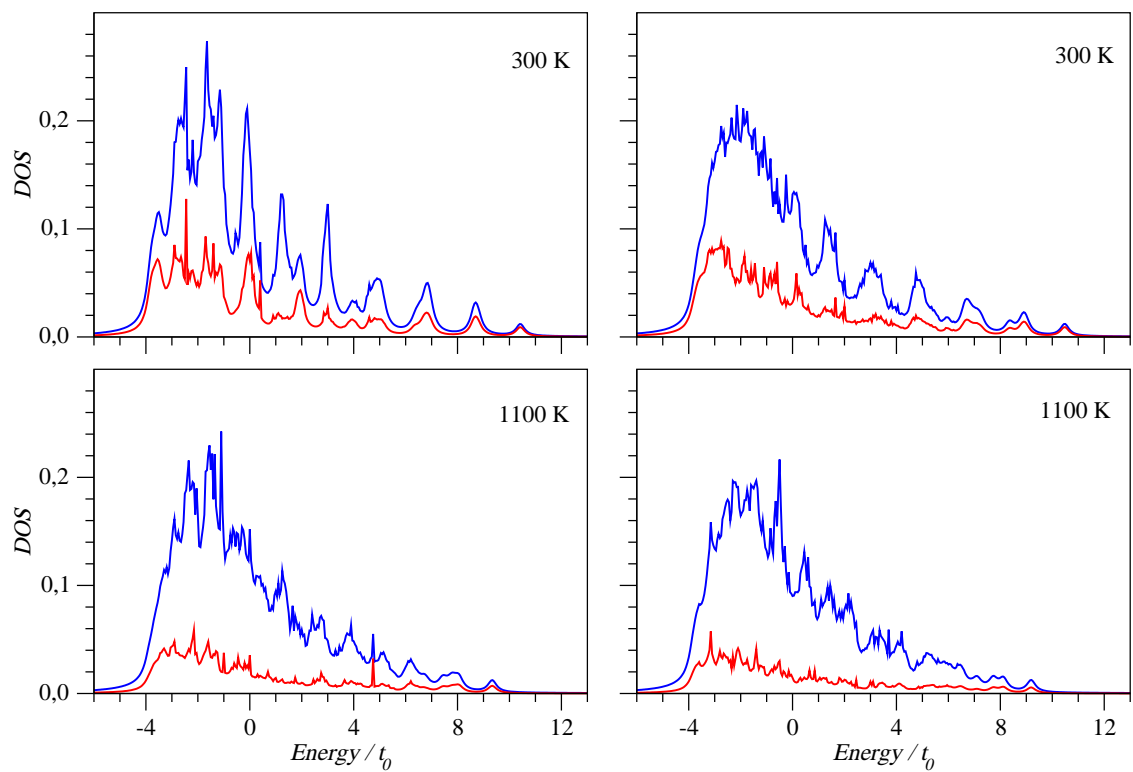}
\caption{total DOS (blue lines), and FDOS (red lines)  
of the Ag${}_{147}$ nanoparticle for two cooling rates: 
$k1=$\qs (a) and b)) and $k2=$\qf (c) and d)). 
The corresponding temperatures are indicated in the panels. }
\label{fig:fig7}
\end{figure}

\subsection{Electronic structure of silver nanoparticles}

To determine the DOS of the silver nanoparticles we consider $t_0=1$eV 
(reference energy), $\varepsilon_0/t_0=0$, $\gamma/t_0=0.2$,
$\Delta\varepsilon/t_0\simeq 0.05$ (energy step), and 75 (20) recursion 
coefficients for the Ag${}_{2869}$ (Ag${}_{147}$) cluster.

Fig. \ref{fig:fig6} shows the DOS of the Ag${}_{2869}$ nanoparticle for 
k1=\qs and k2=\qf$\;$ for both initial (at 1500 K) and final (at 300 K) 
atomic configurations. Comparing the total DOS and the FDOS (fcc-like DOS) 
we can observe that at high temperatures the contribution of FDOS to the
total DOS is not remarkable. This changes completely at 300 K.
For the $k1$ cooling rate, the total DOS at room temperature is slightly
more similar to the DOS of the fcc-solid sample (not shown here), 
especially for the main peak at $\sim-3t_0$, which is not observed in the 
case for $k2$. This should indicate, in agreement with the analysis of
the atomic order (see previous sub-section), that the silver nanoparticle
with a slow cooling rate becomes more ordered (close to its solid 
counterpart) after decreasing the temperature. 
The differences in the total DOS, at room temperatures, for both cooling 
rates are more appreciable in the small nanoparticle (Ag${}_{147}$), 
see Fig.~\ref{fig:fig7}. For this case the FDOS is different from its
fcc-solid counterpart.

\section {Conclusions}

The glass formation and crystallization of a supercooled silver nanodrop, 
4.4 nm in diameter, has been investigated based on the basis of the MD 
simulation with TB-SMA potential. The final structures are highly affected 
by the cooling rates. Ag${}_{2869}$ nanoparticles, obtained in a supercooled liquid 
with cooling rates higher than the critical value 
$k_c = 7.8 \times 10^{12} Ks^{-1}$ are  very unstable metallic glasses, 
and nanoparticles resulting from relatively slow cooling rates are close 
packed crystals (fcc and hcp structures),  as indicated by the CNA technique. 
At the cooling rate of \qs we find a discontinuous structural transition near 
527 K. The fully crystallized nanoparticle is a faceted polycrystal with a 
number of step consisting of one atomic layer. 
The study of the evolution of the DOS with the temperature 
indicates that the electronic structure depends strongly on the size of the
nanoparticle and the cooling rates.

\section*{Acknowledgments}

The authors gratefully acknowledge the financial support of the CSI of the 
National University of San Marcos (Project N 081301061).


\section*{References}

\begin{thebibliography}{99}
\bibitem{Eberh2002} Eberhardt W 2002 \emph{Surf. Sci.} \textbf{500} 242
\bibitem{BalettoRMP} Baletto F and Ferrando R 2005 \emph{Rev. Mod. Phys} \textbf{77} 371
\bibitem{Sidharta} Shrivastava S, Bera T and Roy A 2007 \emph{Nanotechnology} \textbf{18} 225103
\bibitem{GafnerFSS} Gafner Yu, Gafner S and Entel P 2004 \emph{Phys.Solid State} \textbf{46} 1327
\bibitem{Duan2008} Shi D W, He L M, Kong L G, Lin H  and Hong L 2008 \emph{Modelling Simul. Mater. Sci. Eng.} \textbf{16} 025009
\bibitem{YueQi2001} Qi Y, Cagin T, Johnson W L and Goddard W A III 2001 \emph{J. Chem. Phys.} \textbf{115}  385
\bibitem{BalettoCPL} Baletto F, Mottet C and Ferrando R 2002 \emph{Chem. Phys. Lett.} \textbf{354} 82-87
\bibitem{Nam2002} Nam H S, Hwang N M, Yu B D and Yoon J K 2002 \emph{Phys.Rev. Lett.} \textbf{89} 275502
\bibitem{Shim2003} Shim J-H, Lee S-C, Lee B-J, Suh J-Y and Whan Cho Y 2003 \emph{J. Cryst. Growth}  \textbf{250} 558
\bibitem{Chen2004} Chen Y, Bian X, Zhang J, Zhang Y and Wang L 2004 \emph{Modelling Simul. Mater. Sci. Eng.} \textbf{12} 373
\bibitem{Delog2007} Delogu F 2007 \emph{Nanotechnology} \textbf{18} 485710
\bibitem{AtisAg} Atis M, Aktas H and Guvenc Z 2005 \emph{Modelling Simul. Mater. Sci. Eng} \textbf{13} 1411
%
\bibitem{Qi2008} Qi W H, Wang M P, Liu F X, Yin Z M and Huang B Y 2008 \emph{Comput. Mater. Sci.} \textbf{42} 517
%
\bibitem{Tian08} Tian Z-A, Liu R-S, Liu H-R, Zheng C-X, Hou Z-Y and Peng P 2008 \emph{J. Non-Cryst. Solids} \textbf{354} 3705.
%
\bibitem{Baletto2002} Baletto F, Ferrando R, Fortunelli A, Montalenti F and Mottet C 2002 \emph{J. Chem. Phys.} \textbf{16} N9 3856
\bibitem{Reinhard} Reinhard D, Hall B D, Ugarte D and Monot R 1997 \emph{Phys.Rev. B} \textbf{55} 7868
\bibitem{CleriRos} Cleri F and Rosato V 1993 \emph{Phys.Rev. B} \textbf{48} 22
\bibitem{Still} Stillinger F H and Weber T A 1982 \emph{Phys. Rev. A} \textbf{25} 978
\bibitem{Honey} Honeycutt J D and Andersen H C 1987 \emph{J.Phys. Chem.} \textbf{91} 4950
\bibitem{Haydock80} Haydock R, Heine V, Kelly M L and Bullet D W 1980 \emph{Solid State Phys.} \textbf{35} 215
\bibitem{eye} Li J 2003  \emph{Modelling Simul. Mater. Sci. Eng.} \textbf{11} 173
\bibitem{Hui2002} Li H, Wang G,  Zhao J and Bian X 2002 \emph{J. Chem. Phys.} \textbf{116} 24 10809
%
%
	
\end{thebibliography}
\end{document}